\documentclass{ws-procs975x65}

\begin{document}

\title{SIMILARITY SOLUTIONS OF FOKKER-PLANCK EQUATION WITH TIME-DEPENDENT COEFFICIENTS AND FIXED/MOVING BOUNDARIES}

\author{CHOON-LIN HO$^*$ }

\address{Department of Physics, Tamkang University,\\
 Tamsui 25137, Taiwan, R.O.C.\\
$^*$E-mail: hcl@mail.tku.edu.tw}

\begin{abstract}
We consider the solvability of the Fokker-Planck
equation with both time-dependent drift and diffusion coefficients
by means of the similarity method. By the introduction of the
similarity variable, the Fokker-Planck equation is reduced to an
ordinary differential equation.  Adopting the natural requirement
that the probability current density vanishes at the boundary, the
resulting ordinary differential equation turns out to be
integrable, and the probability density function can be given in
closed form. New examples of exactly solvable Fokker-Planck
equations are presented.
\end{abstract}


\bodymatter

%%%%%%%%%%%%%%%%%%%

\section{Introduction}

One of the basic tools which
is widely used for studying the effect of fluctuations in
macroscopic systems is the Fokker-Planck equation (FPE) \cite{RIS:1996}.
This equation has found applications in such diverse areas as
physics, chemistry, hydrology, biology, finance and
others. Because of its broad applicability, it is therefore of
great interest to obtain solutions of the FPE for various physical
situations.

Generally, it is not easy to find analytic solutions of the FPE, except in a few simple cases,
such as linear drift and constant diffusion coefficients. In most cases, one can only solve the equation
approximately, or numerically.  Most of these methods, however, are concerned
only with FPEs with time-independent diffusion and drift
coefficients (for a review of these methods, see eg. Ref. ~\refcite{RIS:1996}).

Solving the FPEs with time-dependent drift and/or diffusion
coefficient is in general an even more difficult task. It is
therefore not surprising that the number of papers on such kind of
FPE is far less than that on the FPE with time-independent
coefficients.  .

One method of solving a differential equation using symmetry property is the so-called
similarity method \cite{BC:1974}.  This method is especially useful when the system under consideration possesses scaling symmetry, i.e., when it has the property of
self-similarity either on different time and/or space scales.  The well-known diffusion equation, being a special case of the FPE, is one such system. 
 
One advantage of the similarity method is that it allows one to
reduce the FPE to an
ordinary differential equation which is generally easier to solve, provided that the FPE
possesses proper scaling property under certain scaling
transformation of the basic variables.  Some interesting exactly solvable cases of such FPE on the real line $x\in (-\infty,\infty)$ and
the half lines $x\in [0,\infty)$ and $x\in (-\infty,0]$  were considered in Ref.~\refcite{LH}.
These domains admit similarity solutions because their boundary points are the fixed points of the scaling transformation considered.
This indicates that similarity solutions are not possible for other finite domains.

However, similarity solutions of  FPE on a finite domain may be possible, if its boundary points scale accordingly. This leads to FPE with moving boundaries.
Examples of such FPEs are presented in Ref.~\refcite{Ho}.

\section{Scaling of Fokker-Planck equation}

We first discuss the scaling form of the FPE.
The general form of the FPE in $(1+1)$-dimension is
\begin{eqnarray}\label{E2.1}
\frac{\partial W(x,t)}{\partial t}=\Big[-\frac{\partial}{\partial x}
D^{(1)}(x,t)+\frac{\partial^2}{\partial x^2}D^{(2)}(x,t)\Big]W(x,t)\;,
\end{eqnarray}
where $W(x,t)$ is the probability distribution function,
$D^{(1)}(x,t)$ is the drift coefficient and $D^{(2)}(x,t)$ the
diffusion coefficient. The drift coefficient represents the
external force acting on the particle, while the diffusion
coefficient accounts for the effect of fluctuation. $W(x,t)$ as a
probability distribution function should be normalized, $i.e.$,
$\int_{\textstyle\mbox{\small{domain}}}W(x,t)\,dx=1$ for $t\geq
0$.

We shall be interested in seeking similarity solutions of the FPE, which are possible if
the FPE possesses certain scaling symmetry.

Consider the scale transformation
\begin{align}\label{E2.2}
&\bar{x}=\varepsilon^a x\;\;\;,\;\;\;\bar{t}=\varepsilon^b t,
\end{align}
where $\varepsilon$, $a$ and $b$ are real parameters. Suppose
under this transformation, the probability density function and
the two coefficients scale as
\begin{align}\label{E2.3}
\bar{W}(\bar{x},\bar{t})=\varepsilon^c W(x,t),~
\bar{D}^{(1)}(\bar{x},\bar{t})=\varepsilon^d
D^{(1)}(x,t), ~\bar{D}^{(2)}(\bar{x},\bar{t})=\varepsilon^e
D^{(2)}(x,t).
\end{align}
Here $c$, $d$ and $e$ are also some real parameters.  It can be checked that
the transformed equation in terms of the new variables has the same functional form as eq.(\ref{E2.1})
if the scaling indices satisfy $b=a-d=2a-e$. In this case, the second order FPE can  be transformed into an
ordinary differential equation which is easier to solve. Such reduction is effected
through a new independent variable $z$ (called similarity variable), which is
certain combination of the old independent variables such that
it is scaling invariant, i.e., no appearance of parameter
$\varepsilon$, as a scaling transformation is performed. Here
the similarity variable $z$ is defined by
\begin{align}\label{E3.1}
z\equiv\frac{x}{t^{\alpha}}\;,\;\;\;\mbox{where}\;\;\;
\alpha=\frac{a}{b}\;\;\;\mbox{and}\;\;\;a\;,b\neq 0\;.
\end{align}
For $a\;,b\neq 0$, one has $\alpha\neq 0\;,\infty$.

The scaling form of the probability density function
can be taken as \cite{LH}
\begin{align}\label{E3.2}
W(x,t)=t^{\alpha\frac{c}{a}}y(z)\;,
\end{align}
where $y(z)$ is a function of $z$. The normalization of the distribution function is
\begin{align}\label{E3.5}
\int_{\mbox{\small{domain}}}\,W(x,t)\,dx=\int_{\mbox{\small{domain}}}\,
\Big[t^{\alpha(1+\frac{c}{a})}\,y(z)\Big]\,dz=1\;.
\end{align}
For the above relation to hold at all $t\geq 0$, the power of $t$
should vanish, and so one must have $c=-a$, and thus
\begin{equation}
W(x,t)=t^{-\alpha}y(z). \label{W-scale}
\end{equation}
Similar consideration leads to the following scaling forms of the
drift  and diffusion coefficients
\begin{align}\label{E3.3}
D^{(1)}(x,t)=t^{\alpha-1}\rho_1(z)\;\;\;,\;\;\;D^{(2)}(x,t)=t^{2\alpha-1}\rho_2(z)\;,
\end{align}
where $\rho_1(z)$ and $\rho_2(z)$ are scale invariant functions of $z$.

With Eqs.~(\ref{E3.1}), (\ref{W-scale}) and (\ref{E3.3}), the FPE is
reduced to
\begin{align}\label{E3.4}
\rho_2(z)\,y''(z)+\Big[2\rho_2'(z)-\rho_1(z)+\alpha z\Big]\,y'(z)
+\Big[\rho_2''(z)-\rho_1'(z)+\alpha \Big]\,y(z)=0\;,
\end{align}
where the prime denotes the derivative with respect to $z$.
It is really interesting to realize  that Eq.~(\ref{E3.4}) is exactly integrable.
Integrating it once, we get
\begin{equation}
\rho_2(z) y^\prime(z) + \left[\rho_2^\prime (z) - \rho_1 (z)+
\alpha z\right] y(z)=C, \label{y1}
\end{equation}
where $C$ is an integration constant.
Solution of Eq.~(\ref{y1}) is
\begin{eqnarray}
y(z)&=&\left(C^\prime+C\int^z dz \frac{e^{-\int^z\,dz f(z)}}{\rho_2(z)}\right)\,\exp\left(\int^zdz f(z)\right),\nonumber\\
f(z)&\equiv& \frac{\rho_1(z)-\rho_2^\prime(z)-\alpha
z}{\rho_2(z)},~~~\rho_2(z)\neq 0,
\label{y-soln}
\end{eqnarray}
where $C^\prime$ is an integration constant.

We shall consider boundaries which are impenetrable.  At such boundaries,
the probability density and the associated probability
current density must vanish.
This in turn implies that $C=0$,  and the probability density function $W(x,t)$ is given by
\begin{eqnarray}
W(x,t)=A t^{-\alpha} \exp\left(\int^z
dz\,f(z)\right)_{z=\frac{x}{t^\alpha}},\label{W-soln}
\end{eqnarray}
where $A$ is the normalization constant.  It is
interesting to see that the similarity solution of the FPE can be
given in such an analytic closed form.  Exact similarity solutions
of the FPE can be obtained as long as $\rho_1(z)$ and $\rho_2(z)$
are such that the function $f(z)$ in Eq.~(\ref{W-soln}) is an
integrable function and the resulted $W(x,t)$ is normalizable.
Equivalently, for any integrable function $f(z)$ such that
$W(x,t)$ is normalizable, if one can find a function $\rho_2(z)$
($\rho_1(z)$ is then determined by $f(z)$ and $\rho_2(z)$), then
one obtains an exactly solvable FPE with similarity solution given
by Eq.~(\ref{W-soln}).  Some interesting cases of such FPE on the real line $x\in (-\infty,\infty)$ and
the half lines $x\in [0,\infty)$ and $x\in (-\infty,0]$  were discussed in Ref.~\refcite{LH}.

%------------------------

\section{An Example with fixed boundaries}
%\section{FPE with $\rho_1(z)=\mu_1z+\mu_2$ and $\rho_2(z)=\mu_3 z$}

As an interesting exactly solvable example, let us consider
a FPE  with $\rho_1(z)=\mu_1z+\mu_2$ and $\rho_2(z)=\mu_3z$. The
corresponding drift and diffusion coefficients are
\begin{align}
D^{(1)}(x,t)=\mu_1\,\frac{x}{t}+\mu_2\,t^{\alpha-1}\;\;\;,\;\;\;
D^{(2)}(x,t)=\mu_3\,x\,t^{\alpha-1}\;.
\end{align}
Eq.~(\ref{W-soln}) is integrable and gives
\begin{align}\label{E4.12}
W(x,t)=\frac{\left|\frac{\alpha-\mu_1}{\mu_3\,t^{\alpha}}\right|^{\frac{\mu_2}{\mu_3}}}
{\Gamma\Big(\displaystyle\frac{\mu_2}{\mu_3}\Big)}\,x^{\frac{\mu_2}{\mu_3}-1}\,
\exp\Big\{-\frac{\alpha-\mu_1}{\mu_3\,t^{\alpha}}\,x\Big\}.
\end{align}
The form of $W(x,t)$ implies that the domain of $x$ is defined
only on half-line.  For definiteness we shall take $x\in
[0,\infty)$.  Normalizability of $W(x,t)$ then requires
\begin{equation}
\frac{\alpha-\mu_1}{\mu_3}>0,~~\frac{\mu_2}{\mu_3}\geq 1.
\end{equation}

\begin{figure}[b]%
\begin{center}
\parbox{2.1in}
{\epsfig{figure=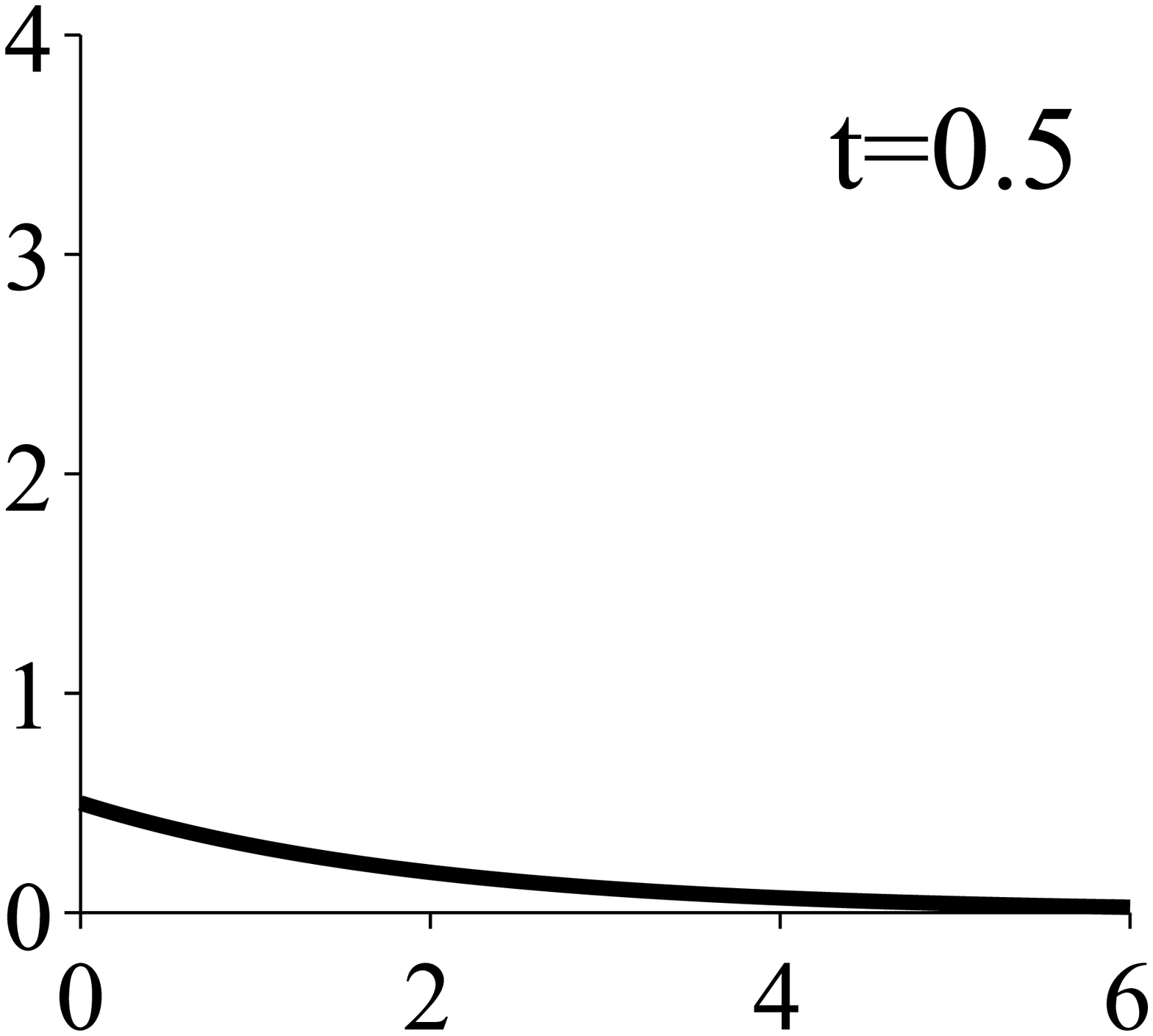,width=2in}}
 %\figsubcap{a}}
 \hspace*{4pt}
  \parbox{2.1in}
{\epsfig{figure=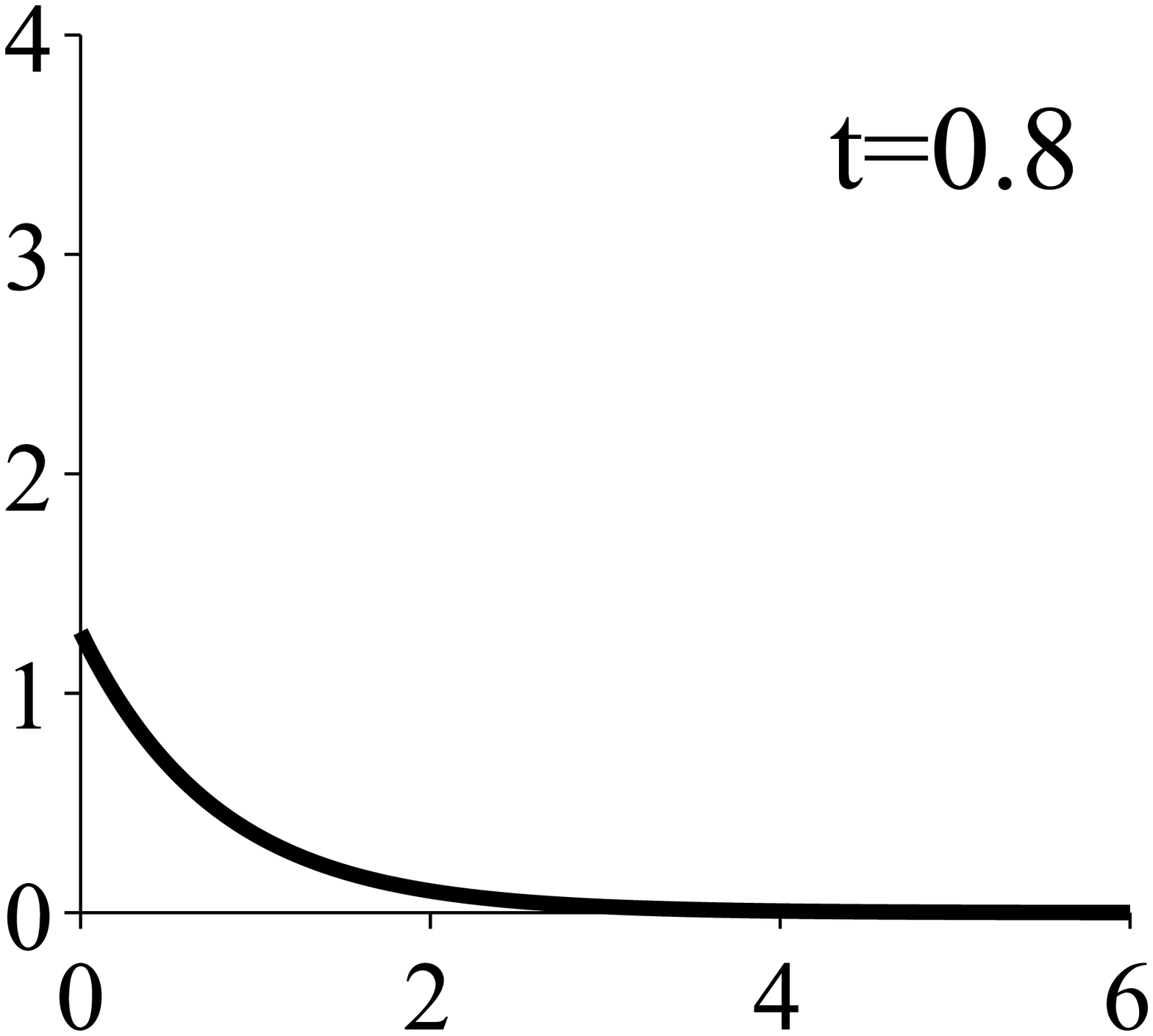,width=2in}}
% \figsubcap{b}}
\\
 \parbox{2.1in}
{\epsfig{figure=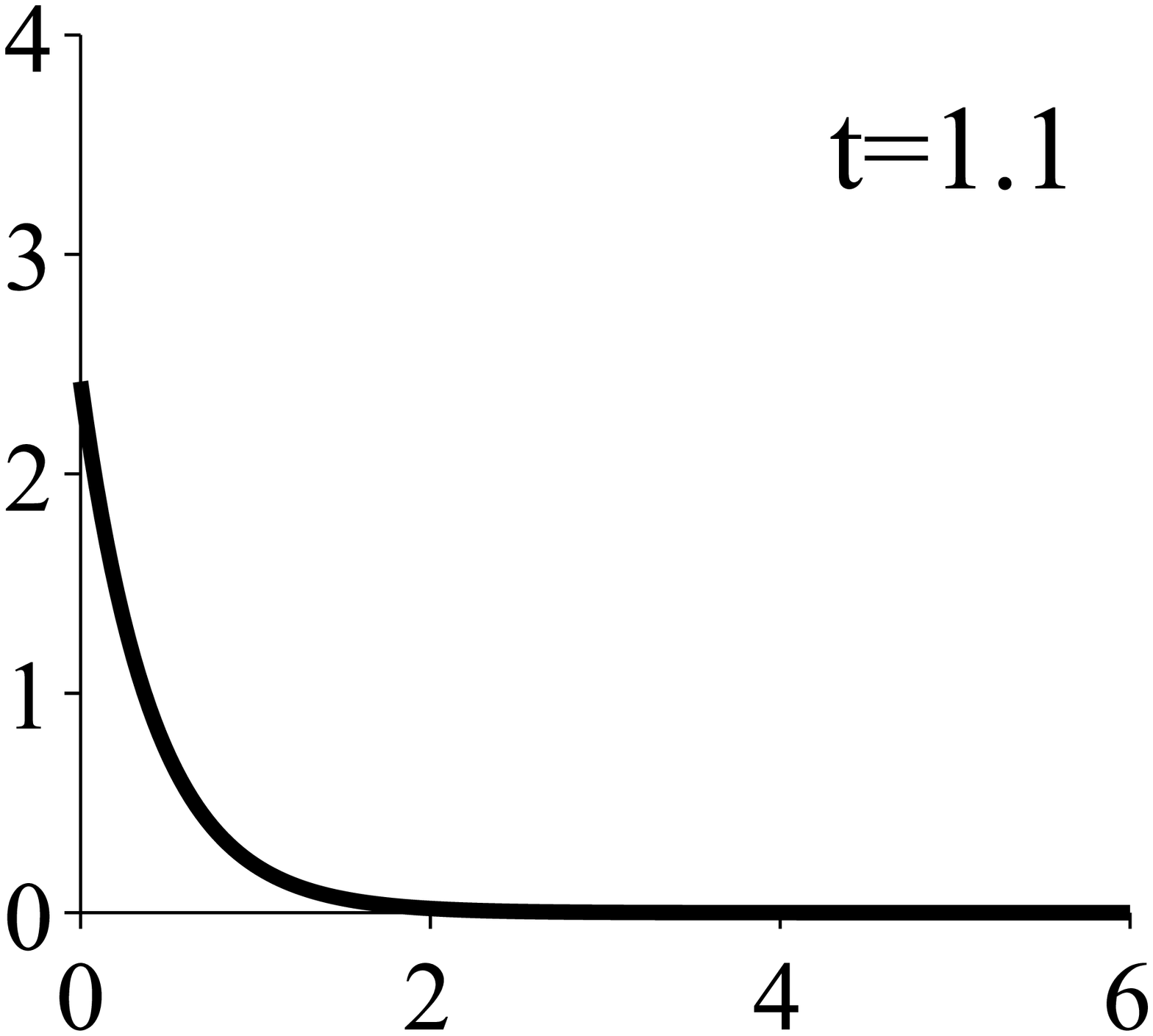,width=2in}}
% \figsubcap{a}}
 \hspace*{4pt}
 \parbox{2.1in}
{\epsfig{figure=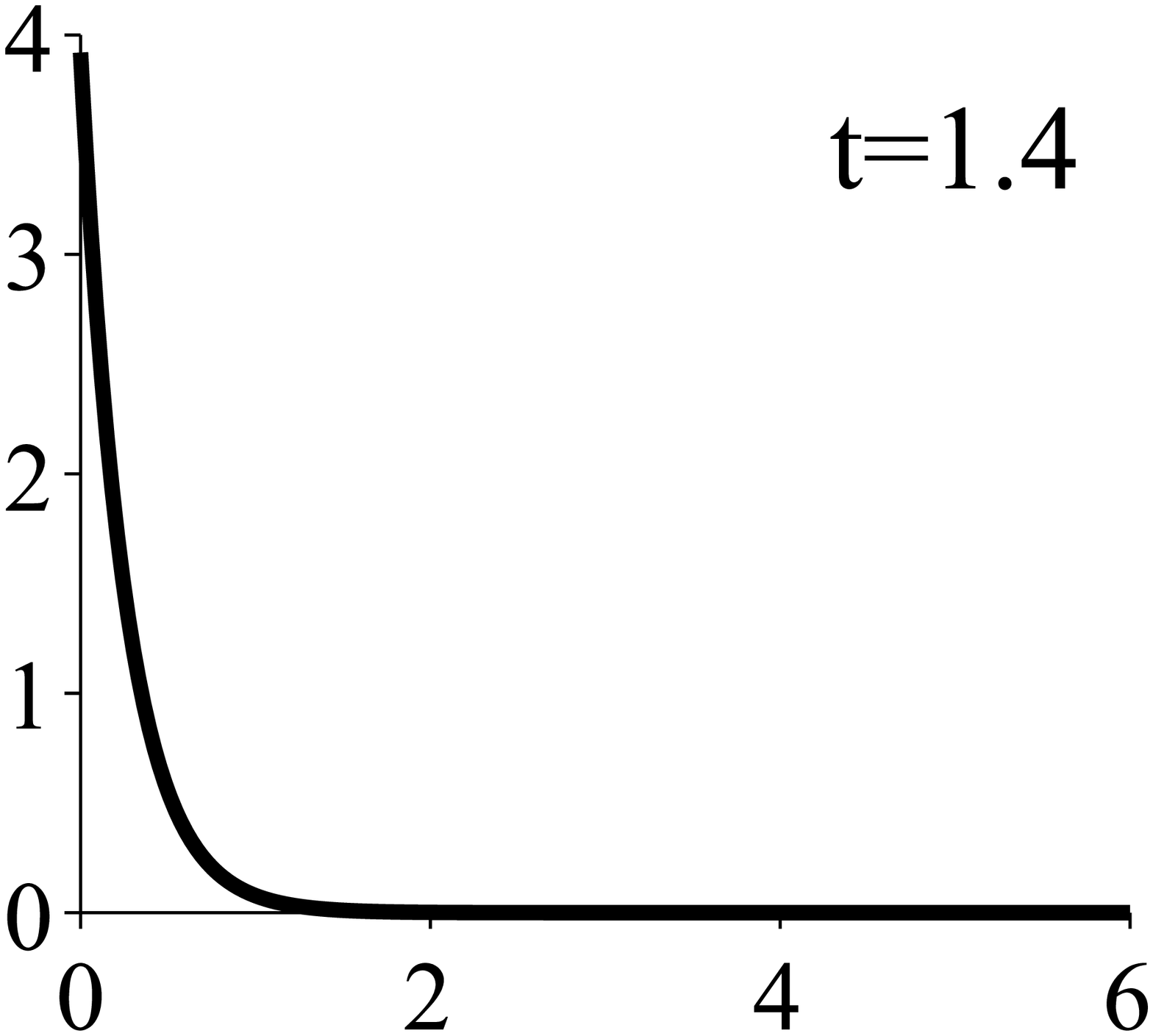,width=2in}}
% \figsubcap{b}}
\caption{Plot of $W(x,t)$ versus $x$ for solution~(\ref{E4.12})
with $\alpha=-2$, $\mu_1=-3$, $\mu_2=\mu_3=1/2$, and time $t=0.5$,
$0.8$, $1.1$, $1.4$.} 
\label{Fig.1}
\end{center}
\end{figure}

Solution (\ref{E4.12}) with $\mu_2=\mu_3$ presents an interesting
stochastic process. In this case, $W(x,t)$ becomes the exponential
function, whose peak is always located at the origin. Its peak
value is $|(\mu_1-\alpha)/(\mu_3\,t^{\alpha})|$, which is
dependent on the parameters $\alpha$, $\mu_1$ $\mu_3$ and time
$t$, and hence is affected by both the drift and diffusion
coefficients.  The peak at $x=0$ is increasing (decreasing) as $t$
increases for $\alpha<0$ ($\alpha>0$).  That means, by an
appropriate choice of the drift and diffusion parameters, one can
have a situation where the probability function is accumulating at
the origin. For such situation, the effect of the drift force is
stronger than that of the diffusion,  causing the distribution to
be pushed toward the origin as time elapses. An example of such
situation is depicted in FIG.~\ref{Fig.1}, which demonstrates
the evolution of solution (\ref{E4.12}) with $\alpha=-2$ and
$\mu_2=\mu_3$.

\section{FPEs with moving boundaries}

We now illustrate the construction of exactly solvable FPEs with moving boundaries.  Only one class of such system is presented here.  Other classes are discussed in Ref.~\refcite{Ho}.

We consider a finite domain $x_1(t)\leq x \leq x_2(t)$ with impenetrable moving boundaries at $x_k(t)~(k=1,2)$.
We want the transformed FPE in $z$-space to be exactly solvable.  The simplest choice is that in the $z$-space the
boundary points of the corresponding domain are static.  This implies $ z_k = x_k(t)/t^\alpha (k=1,2)$ are constants.
Note that the fixed  domains admitting similarity solutions considered  in
Ref.~\refcite{LH} correspond to the choice $z=x(t)/t^\alpha=0,\pm\infty$, which in the $x$-space are just the fixed points of the scaling transformation .

If we choose the function $f(z)$ to have the form
\begin{eqnarray}
f(z)=\frac{a_1}{z-z_1}-\frac{a_2}{z_2-z},~~a_1,\,\,a_2>0, ~z_1\leq z\leq z_2,
\end{eqnarray}
then  $\rho_1(z)$ and $\rho_2(z)$ are given by
\begin{align}
\rho_2 (z)&=(z-z_1)(z_2-z),\\
\rho_1 (z)&=(\alpha-a_1-a_2-2)z+(a_1+1)z_2 + (a_2+1)z_1
\end{align}
for $z_1\leq z\leq z_2$, and $\rho_1 (z),\,\rho_2 (z)=0$ otherwise.
The function $y(z)$ in the physical domain is
\begin{eqnarray}
y(z)=A(z-z_1)^{a_1} (z_2-z)^{a_2}.
\end{eqnarray}
Here the normalisation constant $A$ is given by
\begin{eqnarray}
A=[(z_2-z_1)^{a_1+a_2+1}B(a_1+1,a_2+1)]^{-1},
\end{eqnarray}
where $B(x,y)$ is the Beta function.

The probability density function is
\begin{eqnarray}
W(x,t)=
\left\{
\begin{array}{ll}
 \frac{A}{t^\alpha}\left(\frac{x}{t^\alpha}-z_1\right)^{a_1}\left(z_2-\frac{x}{t^\alpha}\right)^{a_2},&z_1t^\alpha\leq x\leq z_2 t^\alpha\\
0,&{\rm otherwise}
 \end{array}\right..
\end{eqnarray}
There are three subclasses:
\begin{align}
({\rm i})~  z_1,\, z_2>0 ~ (z_1,\,z_2<0);~~~
({\rm ii})~  z_1=0,\, z_2>0 ~ (z_1<0,\, z_2=0);~~~
({\rm iii})~ z_1<0, \, z_2>0.
\end{align}
The situations given in the brackets correspond to mirror images of the corresponding classes with an appropriate change of parameters.
In subclass (ii), $z_1=0$ is a fixed point of the scale transformation, and can be considered as a special case of Case II to be discussed below.
It is found that
for $\alpha>0 (<0) $, the boundaries move away from (toward) the origin (except when the boundary is a fixed point).

\begin{figure}
\psfig{file=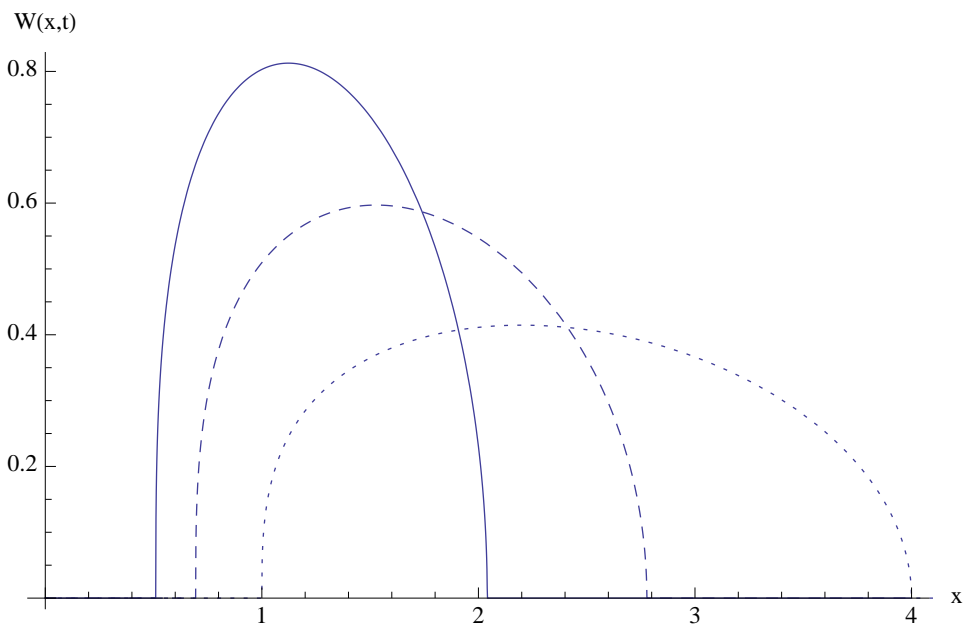,width=5in}
\caption{$W(x,t)$ vs $x$ for Case-I(i) with $\alpha=-2, z_1=1, z_2=4,a_1=1/3$ and $a_2=1/2$ for $t=1.0$ (dotted line), $1.2$ (dashed line) and $1.4$ (solid line). }
\label{Fig.2}
\end{figure}

In Fig.~\ref{Fig.2} we show figures for  subclass (i)  with $\alpha<0$, showing an overall left-moving density function towards the origin. 

%\begin{figure}
 %\centering
 %\includegraphics{FPE-MB-Case-I-i-n.eps}
%\caption{$W(x,t)$ vs $x$ for Case-I(i) with $\alpha=-2, z_1=1, z_2=4,a_1=1/3$ and $a_2=1/2$ for $t=1.0$ (dotted line), $1.2$ (dashed line) and $1.4$ (solid line). }
%\label{Fig.1}
%\end{figure}

\end{document}